\documentclass[fleqn,usenatbib]{mnras}

\usepackage{newtxtext,newtxmath}

\usepackage[T1]{fontenc}

\DeclareRobustCommand{\VAN}[3]{#2}
\let\VANthebibliography\thebibliography
\def\thebibliography{\DeclareRobustCommand{\VAN}[3]{##3}\VANthebibliography}

\usepackage{graphicx}	
\usepackage{amsmath}	

\title[Brown dwarfs in NGC 2264]{The first spectroscopically confirmed brown dwarfs in NGC 2264}

\author[S. Pearson et al.]{
Samuel Pearson$^{1}$,\thanks{E-mail: sp246@st-andrews.ac.uk}
Aleks Scholz$^{1}$,
Paula S Teixeira$^{1}$,
Koraljka Mu\v{z}i\'{c}$^{2}$,
Víctor Almendros-Abad$^{2}$
\\
$^{1}$ SUPA, School of Physics \& Astronomy, University of St Andrews, North Haugh, St Andrews, KY16 9SS, United Kingdom \\
$^{2}$ CENTRA, Faculdade de Ci\^{e}ncias, Universidade de Lisboa, Ed. C8, Campo Grande, 1749-016 Lisboa, Portugal\\
}

\date{Accepted XXX. Received YYY; in original form ZZZ}

\pubyear{2021}

\begin{document}
\label{firstpage}
\pagerange{\pageref{firstpage}--\pageref{lastpage}}
\maketitle

\begin{abstract}
We present spectroscopic follow-up observations of 68 red, faint candidates from our multi-epoch, multi-wavelength, previously published survey of NGC 2264. Using near-infrared spectra from VLT/KMOS, we measure spectral types and extinction for 32 young low-mass sources. We confirm 13 as brown dwarfs in NGC 2264, with spectral types between M6 and M8, corresponding to masses between 0.02 and 0.08\,$M_{\odot}$. These are the first spectroscopically confirmed brown dwarfs in this benchmark cluster. 19 more objects are found to be young M-type stars of NGC 2264 with masses of 0.08 to 0.3$\,M_{\odot}$. 7 of the confirmed brown dwarfs as well as 15 of the M-stars have IR excess caused by a disc. Comparing with isochrones, the typical age of the confirmed brown dwarfs is $<0.5$ to 5\,Myr. More than half of the newly identified brown dwarfs and very low mass stars have ages $<0.5$\,Myr, significantly younger than the bulk of the known cluster population. Based on the success rate of our spectroscopic follow-up, we estimate that NGC 2264 hosts 200-600 brown dwarfs in total (in the given mass range). This would correspond to a star-to-brown dwarf ratio between 2.5:1 and 7.5:1. We determine the slope of the substellar mass function as $\alpha = 0.43^{+0.41}_{-0.56}$, these values are consistent with those measured for other young clusters. This points to a uniform substellar mass function across all star forming environments.
\end{abstract}




\section{Introduction}
As one of the youngest and richest star forming regions accessible to deep surveys, NGC 2264 provides a perfect laboratory to study star and brown dwarf formation. Due to its large size of $\sim1500$ stellar members \citep{Teixeira2012, Rapson2014, Venuti2018}, relative proximity of 719 $\pm$ 16 pc \citep{2264dist} and low foreground extinction, NGC 2264 has been extensively studied for over 60 years \citep{Herbig1954} and will remain a principle region for star formation studies for the foreseeable future.

NGC 2264 is hierarchically structured with several subclusters of members.
Separate populations in the cluster exhibit a range of ages (1-5 Myr), with evidence for sequential star formation starting in the north and spreading to the south \citep{Sung2010, Venuti2018, nony2020}. The northern population surrounds the O-B-B triple system S Mon \citep{mon15}. The younger and more embedded southern population is located close to the tip of the Cone Nebula. NGC 2264 is associated with the Monoceros OB1 molecular cloud complex, which covers over 2 square degrees on the sky.

Apart from the Orion Nebula Cluster (ONC), NGC 2264 is arguably the most well studied, massive star forming cluster within 1\,kpc younger than 5\,Myr. The ONC harbours multiple populations of young stars and it is superimposed onto a bright, extensive emission nebula, which complicates survey work. NGC 2264 is significantly simpler in its architecture.
The current consensus view of this cluster has come from several decades of surveys, however, to our knowledge no brown dwarfs have been spectroscopically confirmed in this benchmark cluster. 

The origin of brown dwarfs and the substellar mass function have long been subjects of debate. The prerequisite for making progress on these issues is to have large and well characterised samples of young brown dwarfs, in diverse star forming regions. NGC 2264 is a diverse star forming region that has the potential to provide this large sample, but currently lacks the spectroscopic observations needed for detailed characterisation. Several purely photometric surveys have probed the substellar domain, including \citet{Lamm2004, Kendall2005, Sung2010}. We have recently published the first photometric study identifying brown dwarf candidates down to estimated masses of 0.02$\,M_{\odot}$, using multi-epoch, multiwavelength data from CTIO/Blanco, KPNO/FLAMINGOS, and Spitzer/IRAC \citep{Pearson20}. In this paper we present spectra for the first confirmed brown dwarfs in NGC 2264 and measure the overall success rate of our candidates. This will allow us to quantify the substellar inventory of this cluster, building an essential sample of young brown dwarfs, as well as lay the groundwork for extending the mass function to the substellar domain.

This is part of an ongoing effort to identify, quantify and characterise the brown dwarf population in a diverse group of young clusters. In the long-term project SONYC, we have targeted a sample of nearby star forming regions, including NGC 1333 \citep{Scholz2012}, Chamaeleon-I, Lupus, \citep{muzic2015}, and Rho Ophiuchus \citep{muzic2011}. More recently we have extended this type of survey to more distant, more massive, and more diverse regions \citep{kora2017, kora2019}. So far we find that the stellar density and the presence or absence of OB stars does not have a significant effect on brown dwarf numbers (relative to stars), a result that provides constraints on formation scenarios for brown dwarfs \citep{kora2019}. NGC 2264 is relevant in this context, since it includes two sub-clusters, one dominated by an O-star, the other one without. 

In Section 2 we outline the target selection criteria, observations and data reduction. In Section 3 we determine the spectral types and extinctions for our sources using template fitting and spectral indices. In Section 4 we discuss the properties of the confirmed brown dwarfs. Finally, in Section 5 we estimate the total number of brown dwarfs in this cluster and derive the substellar mass function. We present our conclusions in Section 6.


\section{Observations and data reduction}

\subsection{Target selection}

The spectroscopy targets were selected from the candidates presented in \citet{Pearson20}. In that study, we used optical, near and mid-infrared photometry, time series data, extinction maps and Gaia kinematics to construct a catalogue of 902 faint red sources with indicators of youth for NGC 2264. Within this catalogue 429 sources were selected as brown dwarf candidates based on their infrared colours. We have obtained spectra for 68 targets. The number of target sources belonging to each of the catalogues established in \citet{Pearson20} is shown in Fig. \ref{fig:TarBar}, a brief overview of these catalogues is given below.

VAR and HIGHVAR contain sources that have been identified as variable and highly variable. DISCS contain sources that have infrared excess indicative of a disc. KIN contains sources with Gaia parallax and proper motions consistent with the locus of NGC 2264. HIGHEX contains sources located in regions of high extinction. ROT contains sources where the rotation period has been identified by fitting the light curve with a sine curve. BDC originates from the other samples mentioned above and contains the sources identified as brown dwarf candidates by their IR colours. 


\begin{figure}
	\centering
  	\includegraphics[width=.45\textwidth]{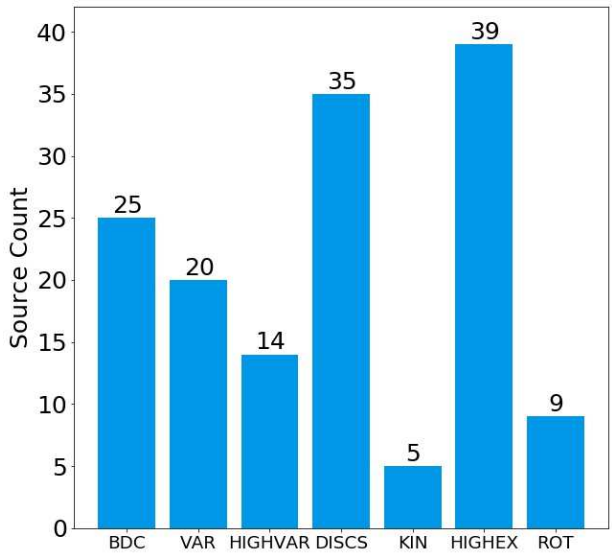}
	\caption{Bar chart showing the number of KMOS target sources in each of the catalogues defined in \citet{Pearson20}. Most sources are in multiple catalogues.}
	\label{fig:TarBar}
\end{figure}

\begin{figure}
	\centering
  	\includegraphics[width=.45\textwidth]{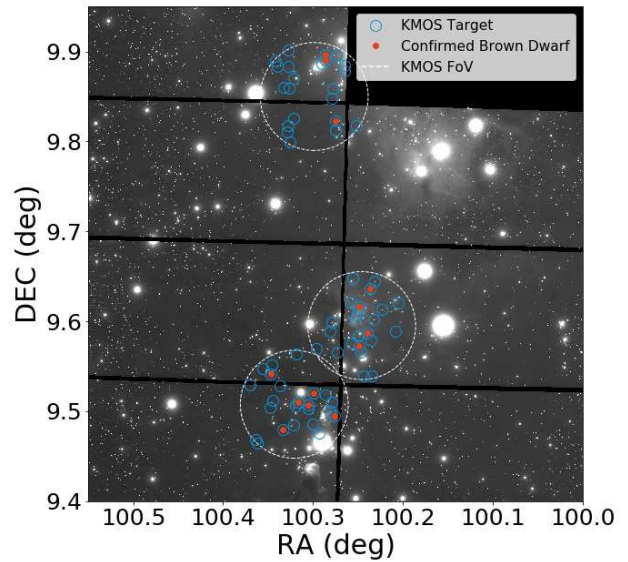}
	\caption{The spatial distribution of the three KMOS fields and the targets for spectroscopy, over-plotted on an I-band image of NGC 2264. The confirmed brown dwarfs are marked in red.}
	\label{fig:TarSpatial}
\end{figure}

\begin{figure*}
	\centering
  	\includegraphics[width=.9\textwidth]{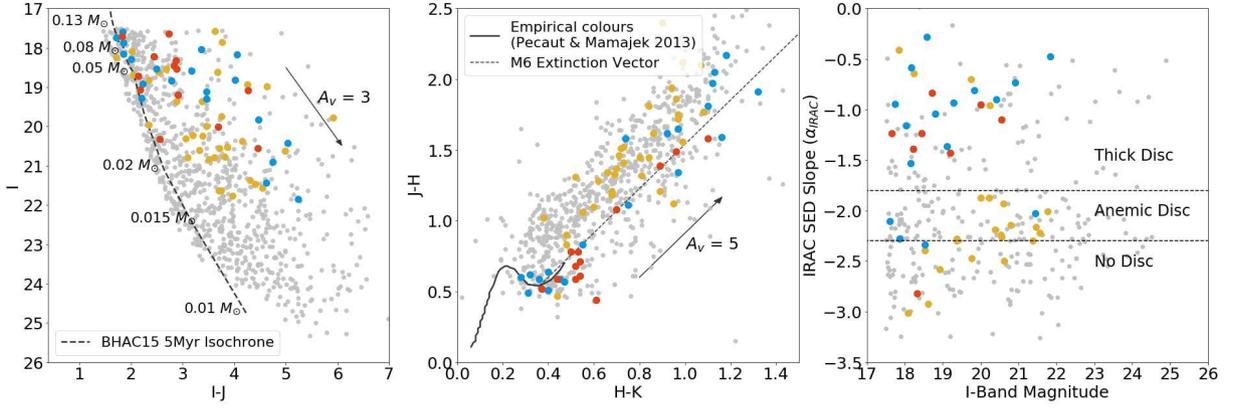}
	\caption{\textit{Left:} I, I-J colour magnitude diagram  overplotted with a 5 Myr BHAC15 isochrone \citep{Baraffe}. The confirmed brown dwarfs ($\ge$M6) are shown in red, the early-mid M stars (M2-M5.5) in blue, the contaminants in yellow and the full Pearson 2020 catalogue in grey. \textit{Middle:} JH, HK colour-colour plot. The solid black line shows the empirical colours for young objects taken from \citet{PandM}, the dashed black line shows the reddened M6 extinction vector. \textit{Right:} The slope of the Spitzer/IRAC SED $\alpha_{IRAC}$ vs. I-band magnitude. The cuts used to differentiate between disc bearing and discless sources are taken from \citet{Lada_2006}. In all three panels, the grey dots show the full catalogue of 902 objects identified in \citet{Pearson20} as being faint, red and having an indication of youth.}
	\label{fig:3pan}
\end{figure*}

\begin{figure}
	\centering
  	\includegraphics[width=.45\textwidth]{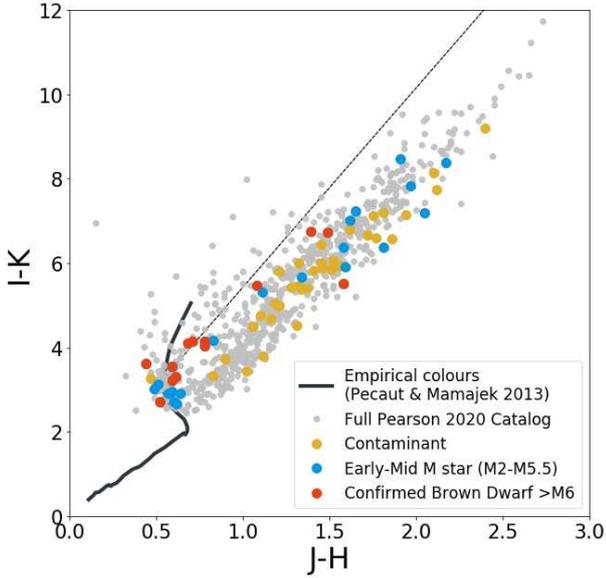}
	\caption{(I-K, J-H) colour-colour plot. The solid black line shows the empirical colours for young objects taken from \citet{PandM}, the dashed black line shows the reddened M6 extinction vector.} 
	\label{fig:IKJH}
\end{figure}

The 68 targets are split across three fields, as shown in Fig. \ref{fig:TarSpatial}. The targets range in brightness from 17.5 - 21 in I. As can be seen from the I, I-J colour magnitude plot in Fig. \ref{fig:3pan} (left), the targets form a subsample consisting of mainly the brighter catalogue sources, the full catalogue extends to $\sim$25 in I. There is a significant range of extinctions, this is illustrated by the large spread of datapoints in the JH, HK colour-colour plot, see Fig. \ref{fig:3pan} (middle). A significant proportion of the targets (34/68) have infrared excess indicative of a disc, this is illustrated in the $\alpha_{IRAC}$ slope vs I-band magnitude plot in Fig. \ref{fig:3pan} (right). Our sample is biased towards disc bearing sources as mid-infrared excess was used as one of the primary selection criteria. In Fig. \ref{fig:3pan} we also show the sources confirmed as young brown dwarfs as red datapoints.

\subsection{Observations}

Spectroscopy of the candidates was performed using the K-band Multi Object Spectrograph (KMOS; \citet{kmos}) at the ESO’s Very Large Telescope (VLT), under the program number 0104.C-0105(A). KMOS employs 24 configurable integral field units (IFUs) positioned within a patrol field $7.2'$ in diameter. Each IFU has a $2.8 \times 2.8''$ field of view sampled at $0.2''$. The positions of the IFUs are allocated in advance by the observer in a configuration file.

The data were obtained in service mode spread across nine observing blocks (OBs), each one hour duration, between November and December 2019. The observations were split over three fields, with 3 hours of observing per field -- 2 hours in the YJ band ($1.0\mu m - 1.3\mu m$,  $R\sim3400$) and 1 hour in the HK band ($1.5\mu m - 2.4\mu m$, $R \sim 1800$). The OBs were split into 6 x 260 second exposures for the YJ band \& 10 x 120 second exposures for the HK band. The series of exposures followed a dither pattern and the sky was observed for background subtraction every two science exposures. 

\subsection{Data reduction}

Science and standard star observations were calibrated and reduced using the KMOS: SPARK pipeline \citep{spark} and the ESOREFLEX workflow \citep{esoReflex}. The master calibration files were created following the standard dark subtraction, flat-field, wavelength calibration and illumination correction procedure. KMOS generates its own internal flatfields using 2 lamps mounted in an integrating sphere outside the instrument. KMOS also has internal argon and neon lamps, used for wavelength calibration. The telluric correction and atmospheric model fitting was performed using Molecfit \citep{Molecfit1, Molecfit2}. KMOS has 3 spectrographs and 3 detectors, with 8 IFU arms being fed to each single spectrograph. A telluric star is observed using one IFU for each of the 3 spectrographs. This telluric spectrum is then used to perform the telluric correction for all sources in that spectrograph. 

The minimum signal-to-noise for the final science spectra, averaged across each band, was 7 in the YJ band and 4 in the HK band. The median signal-to-noise across all the spectra was 39 in the YJ band and 37 in the HK band. We note that the HK band data for the third field has lower signal-to-noise due to clouds affecting parts of the observation.

\section{Spectral Analysis}

\subsection{Template fitting}
We estimated the spectral types and extinctions of our sources by fitting them with young ($\lesssim10$ Myr) low-mass spectral standards ranging from M0 to L7, taken from \citet{luhman2017}. As part of their survey, \citet{luhman2017} used gravity sensitive absorption features ($Na I,\ K I,\ H_2O$) to identify these sources as young, low surface gravity objects. The spectral standards increase in increments of 0.5 subtypes for M0-L0, with three additional spectral standards of L2, L4 \& L7 for the later spectral types. The templates were reddened to $A_V$ values ranging from 0 to 20, in steps of 0.5 mag, using the reddening law from \citet{mathis} with $R_V = 3.1$.

\begin{figure*}
	\centering
  	\includegraphics[width=.9\textwidth]{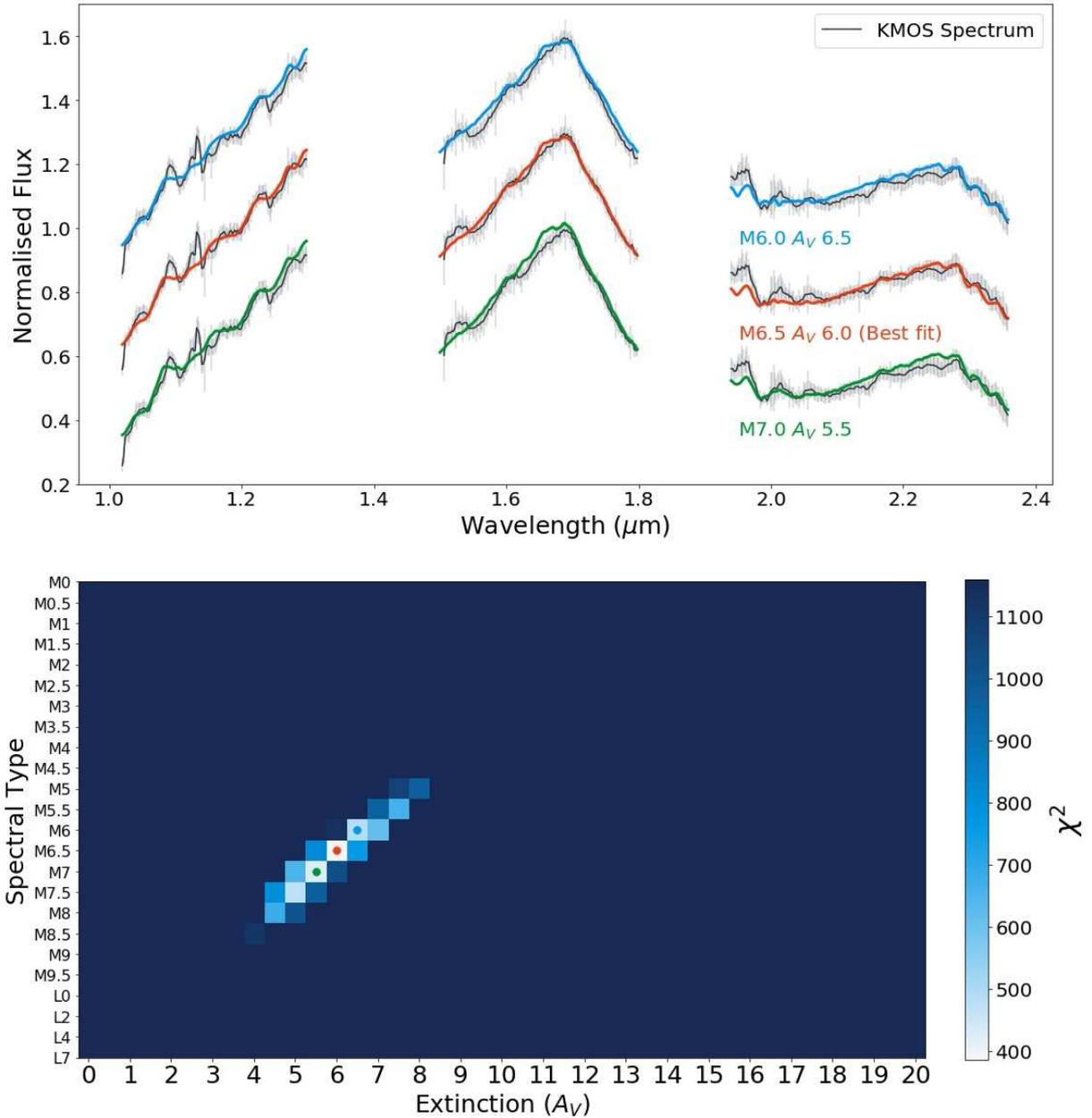}
	\caption{\textit{Top:} The KMOS spectra for brown dwarf (id number 12) is shown in black and over plotted with 3 different spectral templates. The best fitting reddened spectral template (M6.5 $A_V = 6$) is plotted in red. The two second best fits (M6 $A_V = 6.5$  \& M7 $A_V = 5.5$) are plotted in blue and green respectively. \textit{Bottom:} The $\chi^2$ map for a brown dwarf (id number 12). Lighter colours indicate a good fit for both spectral type and extinction. The smallest value for $\chi^2$, marked with a red dot, was taken as the best fit. The green and blue dots indicate the two closest matches after the best fit and correspond to the spectral type and extinction values plotted in green \& blue in the top panel.}
	\label{fig:BestFit}
\end{figure*}

The KMOS spectra from both the YJ and HK bands were smoothed using the \textsc{Astropy} package \textsc{Specutils} \citep{astropy:2013, astropy:2018}. They were then re-sampled to match the resolution of the spectral templates and normalised. For each KMOS source, the $\chi^2$ goodness of fit was calculated for each spectral standard and for each extinction value. This produced a $24\times 41$ grid of $\chi^2$ values in the spectral type vs. $A_V$ parameter space, for each target. An example of these maps is shown in Figure \ref{fig:BestFit}. The $\chi^2$ goodness of fit is defined in equation (1), where \textit{O} is the observed flux at wavelength \textit{i} with variance $\sigma_i^2$, and \textit{T} is the corresponding spectral template flux.

\begin{equation}
    \chi^2 = \sum_i \frac{(O_i - T_i)^2}{\sigma_i^2}
\end{equation}

In the fit, we excluded the noisy $1.8-1.9 \mu m$ window in the HK band, where the signal-to-noise level is very low due to atmospheric extinction. The errors used in the $\chi^2$ calculation were generated by the KMOS reduction pipeline. The minimum $\chi^2$ value was used to identify the best fitting spectral template and extinction for each source. The uncertainty in the fit is taken to be 1 mag in $A_V$, and 1 sub-type in spectral type, as it was found that parameters that are 1-2 grid steps away from the best fit in the spectral type vs. $A_V$ parameter space still produced an acceptable fit within the errorbars of the KMOS data (see Fig. \ref{fig:BestFit}). 

For 33 out of 68 objects, we find a robust fit which is also visually convincing, with a clear structure in the H-band resembling the expected spectral shape of M dwarfs. 13 of those have derived spectral types of M6 or later, \citet{Baraffe} suggest that the hydrogen-burning mass limit occurs near M6 ($T_{\mathrm{eff}} \sim 2925K$) at ages of $\leq10$ Myr. That means, these 13 sources are likely to be young brown dwarfs according to their spectral type. The remaining from the 20 with the spectral signature of M dwarfs are likely young very low mass stars. Through analysis of gravity sensitive features (Sect 3.3), we confirm the youth of all 13 of the brown dwarfs and 19/20 of the low mass stars. We find that 1 of the low mas stars (source id 1), is likely a field dwarf. The other 35 objects in our sample have spectra that are well approximated by reddened blackbodies, but give a poor fit when comparing with the young spectral templates. These will be either embedded young stars with spectral types significantly earlier than M0, or reddened background objects. As these sources are very faint, they are either not detected by Gaia or have errors for their Gaia kinematics that are too large to draw useful conclusions about their cluster membership.

In Table \ref{tab:Big} we report the spectral type and extinction derived using this method. 

\begin{figure}
	\centering
  	\includegraphics[width=.45\textwidth]{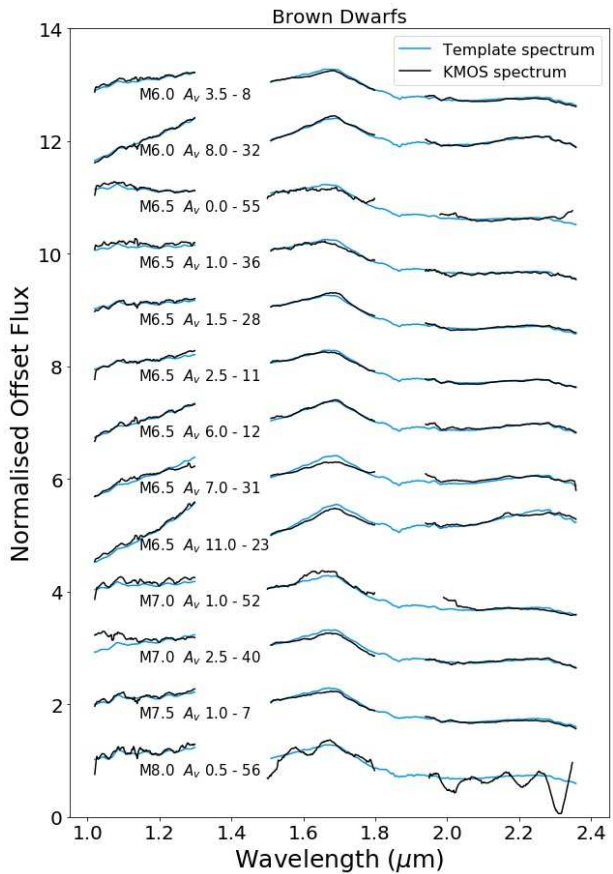}
	\caption{The KMOS spectra and best fitting reddened template for the 13 sources that have been identified as brown dwarfs ($\geq M6$). Each spectrum is annotated with its best fitting spectral type and extinction, as well as its id number.}
	\label{fig:BD}
\end{figure}

\begin{figure*}
	\centering
  	\includegraphics[width=.9\textwidth]{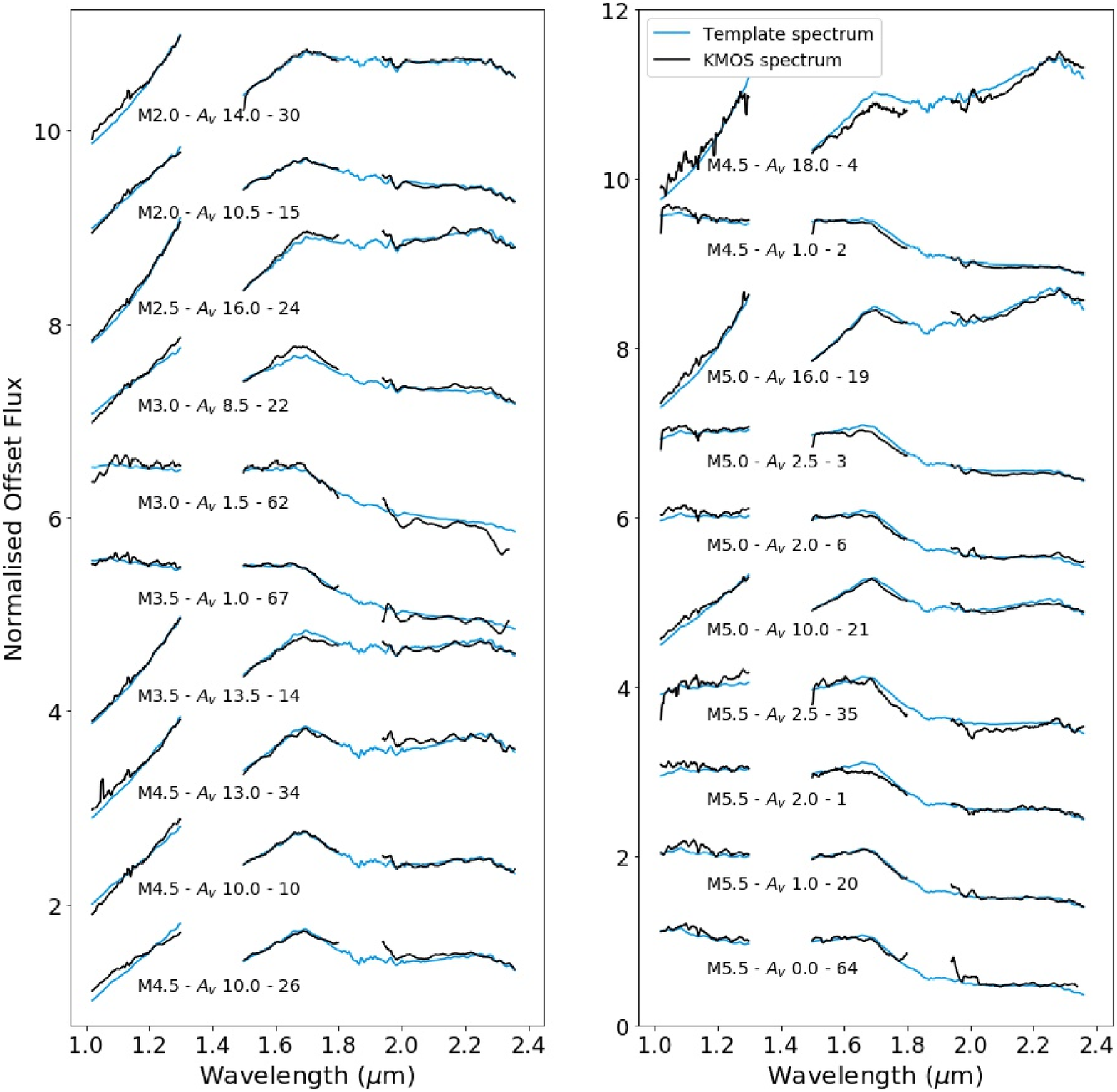}
	\caption{The KMOS spectra and best fitting reddened template for the 20 sources that have been identified as low mas stars (M2-M5.5). Each spectrum is annotated with its best fitting spectral type and extinction, as well as its id number.}
	\label{fig:LMS}
\end{figure*}


\subsection{Spectral Indices}

To assess the validity of the spectral types derived from fitting, we calculate spectral types with an independent, simpler method, using spectral indices calibrated in the literature. Two spectral indices from \citet{weights2009} were used, they are defined in equations (2) and (3). These indices were selected as they use wavelengths covered by the KMOS HK spectra. 

\begin{equation}
    QH = \frac{F_\lambda(1.562\mu m)}{F_\lambda(1.665\mu m)}\left[\frac{F_\lambda(1.740\mu m)}{F_\lambda(1.664\mu m)}\right]^{1.581}
\end{equation}

\begin{equation}
    QK = \frac{F_\lambda(2.050\mu m)}{F_\lambda(2.192\mu m)}\left[\frac{F_\lambda(2.340\mu m)}{F_\lambda(2.192\mu m)}\right]^{1.140}
\end{equation}

\begin{figure}
	\centering
  	\includegraphics[width=.45\textwidth]{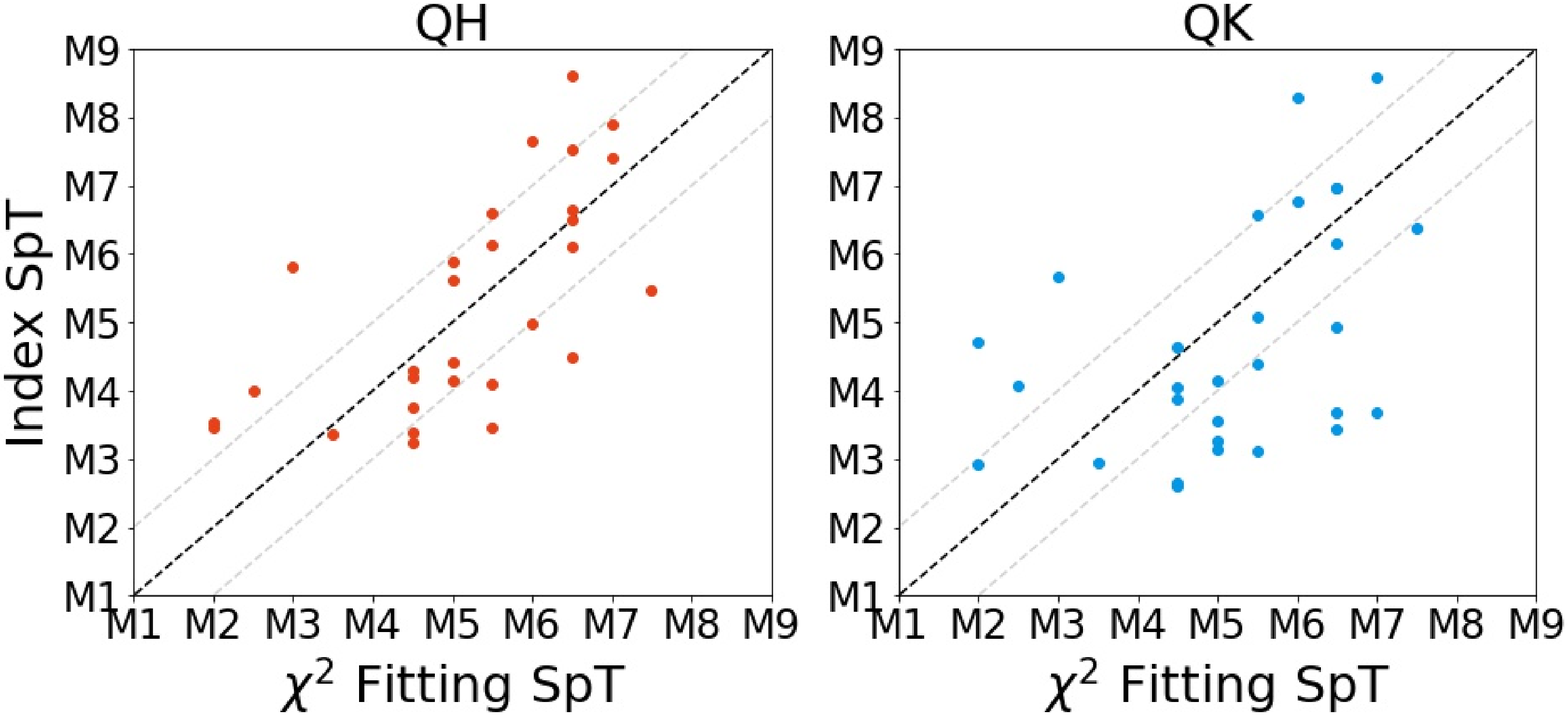}
	\caption{Comparison between the spectral type determined through template fitting and from the spectral indices. The result from the QH index is shown by the red dots, from the QK index by the blue dots. The dashed black line marks an exact agreement between the two methods, the grey lines show the range of typical errors. The median difference between the spectral type from the two indices and the spectral type from the template fitting is 1.03 and 1.43 for the QH and QK indices, respectively.}
	\label{fig:Indices}
\end{figure}

The QH and QK indices are a reddening independent measurements of water absorption. The $F_\lambda$ flux value is determined by the median flux in a $0.02\mu m$ wide interval centered on the specified wavelength. We calculated the QH and QK indices for 29/33 sources identified by spectral fitting as early-mid M stars or brown dwarfs. We were unable to calculate useful values for these indices for 4 of the 33 sources do to the low signal-to-noise for field 3 in the HK band observations. Fig. \ref{fig:Indices} shows a comparison of the spectral types derived from the Weights spectral indices and the chi square fitting. In Table \ref{tab:Indices} we report the spectral type from indices next to the one derived from fitting.


The median disagreement between the spectral type from the two indices and the spectral type from the template fitting is 1.03 and 1.43 for the QH and QK indices respectively. This is equivalent to $\pm 2$ grid steps and agrees with the visually determined uncertainty from the spectral fitting. This provides assurance that the spectral types derived from the fitting are comparable with those estimated from indices, within the given uncertainties. In the following we will only use the spectral types from fitting for further analysis. 

\begin{table}
\label{tab:Indices}
\scriptsize
\caption{A comparison of the spectral types determined by template fitting and the QH \& QK spectral indices for 29 early-mid M stars and brown dwarfs. In this table M0 corresponds to $0.0$ and M9 to 9.0.}
\begin{tabular}{cccc}
\hline
id & $SpT_{fit}$ & $SpT_{QH}$ & $SpT_{QK}$  \\
\hline
1  & 5.5        & 3.47  & 4.39  \\
2  & 4.5        & 4.29  & 4.05  \\
3  & 5.0        & 4.41  & 3.25  \\
4  & 4.5        & 3.24  & 2.65  \\
6  & 5.0        & 4.15  & 4.15  \\
7  & 7.5        & 5.48  & 6.37  \\
8  & 6.0        & 4.97  & 6.78  \\
10 & 4.5        & 4.2   & 4.65  \\
11 & 6.5        & 7.53  & 6.97  \\
12 & 6.5        & 6.1   & 3.69  \\
14 & 3.5        & 3.36  & 2.96  \\
15 & 2.0        & 3.53  & 2.93  \\
19 & 5.0        & 5.61  & 3.13  \\
20 & 5.5        & 6.61  & 6.57  \\
21 & 5.0        & 5.89  & 3.57  \\
22 & 3.0        & 5.83  & 5.66  \\
23 & 6.5        & 6.64  & 4.94  \\
24 & 2.5        & 4.0   & 4.07  \\
26 & 4.5        & 3.75  & 3.88  \\
28 & 6.5        & 8.62  & 6.15  \\
30 & 2.0        & 3.45  & 4.71  \\
31 & 6.5        & 4.5   & 3.44  \\
32 & 6.0        & 7.65  & 8.28  \\
34 & 4.5        & 3.4   & 2.6   \\
35 & 5.5        & 6.13  & 3.11  \\
36 & 6.5        & 6.5   & 6.98  \\
40 & 7.0        & 7.4   & 8.59  \\
52 & 7.0        & 7.9   & 3.67  \\
64 & 5.5        & 4.11  & 5.08  \\
\hline
\end{tabular}
\end{table}

\subsection{Gravity-sensitive features}
An additional argument for youth and cluster membership of our sources is the presence of gravity-sensitive features in their spectrum. There are several spectroscopic features in the wavelength range of the KMOS spectra that indicate low surface gravity, such as the KI doublets at 1.18 and 1.24$\,\mu m$, FeH absorption bands at 1.2 and 1.24$\,\mu m$ and the H-band broadband shape \citep{lucas2001, Gorlova2003, McGovern2004, Allers2007, Scholz2012, Allers2013}.

\begin{figure*}
	\centering
  	\includegraphics[width=.9\textwidth]{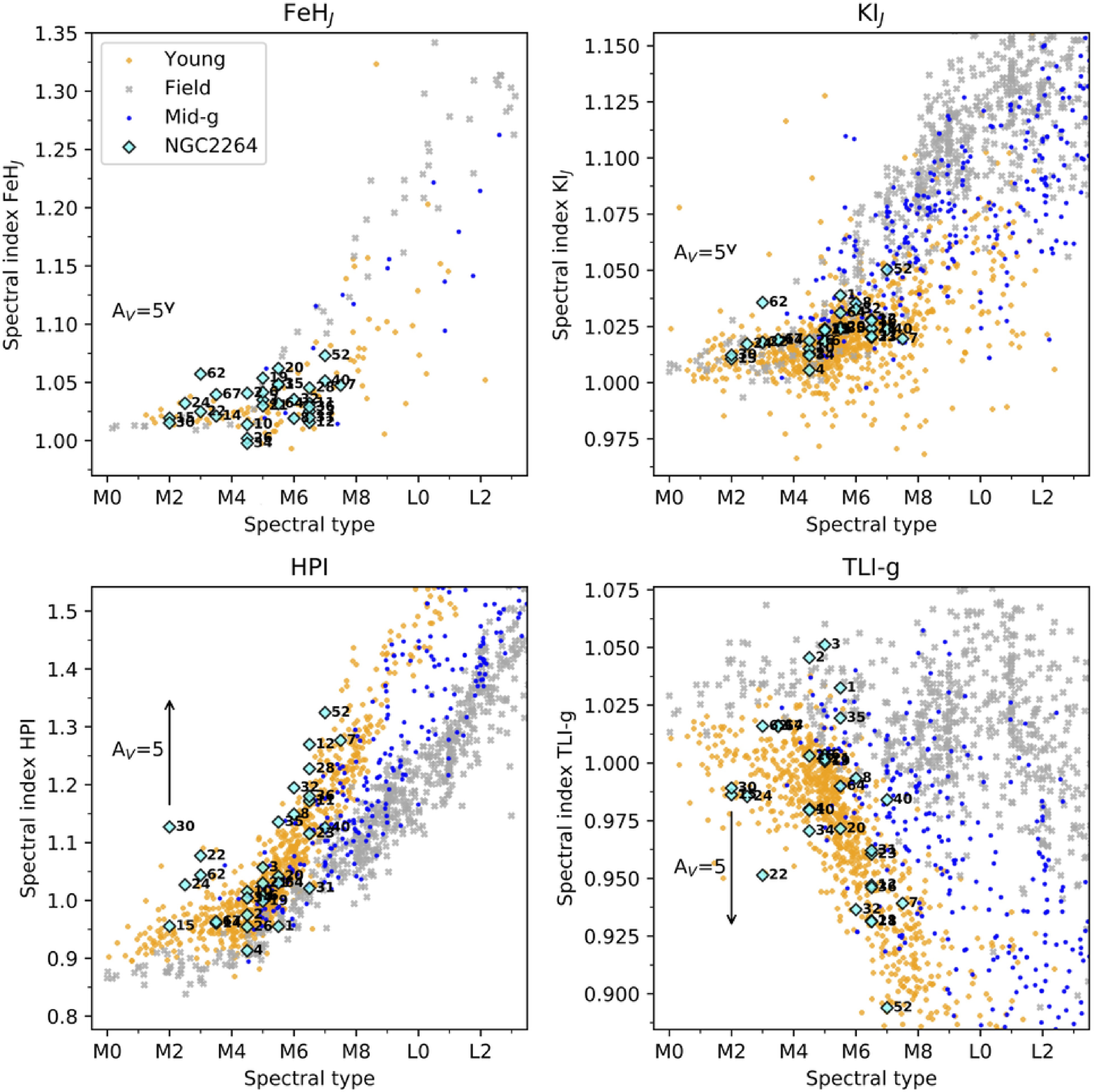}
	\caption{Gravity sensitive indices compared with the spectral type derived in section 3.2. The NGC 2264 objects are shown as cyan diamonds. The rest of the points come from the sample of near infrared dwarf spectra presented in Almendros-Abad et al. (in prep.), and they are colour coded by their gravity class: orange diamonds represent the young, grey crosses the field and blue circles the mid-gravity class objects. A 5 mag reddening vector is also shown.}
	\label{fig:Gravity}
\end{figure*}

In this section we analyse these gravity sensitive features for the 33 objects identified as being low mass (SpT $>$M2). We use the sample of $\sim$2000 near infrared spectra of low mass stars and brown dwarfs presented in Almendros-Abad et al. (in prep.). This sample is composed of all the available (to their best knowledge) near infrared spectra of dwarfs with spectral types between M0-L3. The sample comes from online spectral libraries (e.g the SpeX Prism Library, the Montreal Spectral Library) and individual datasets presented in different works (e.g \citet{Manara2013, Luhman2016, Muirhead2014}). The SpT and extinction of the entire sample was derived in a uniform way by comparing them with spectral templates in a similar manner as presented in Section 3.1 of the present work. This sample, hence forth called DWARFSPEC, is divided into three gravity classes: young (ages up to 10 Myr), mid-gravity (ages from 10 Myr) and field class. By comparing our data to the DWARFSPEC sample for different gravity-sensitive spectral indices we can explore to which gravity class each object of our dataset belongs to. We derive the FeH$_J$, KI$_J$ \citep{Allers2013}, HPI \citep{Scholz2012} and TLI-g (Almendros-Abad et al. in prep.) spectral indices for the 33 objects and plot them together with the DWARFSPEC sources (see Fig. \ref{fig:Gravity}). The DWARFSPEC objects are colour coded by their gravity class, and the KMOS spectra presented in this work are shown as cyan diamonds. We also present a 5 mag reddening vector showing that only the HPI and TLI-g indices are sensitive to extinction. 

The FeH$_J$ and KI$_J$ indices measure the depth of the the 1.20$\mu m$ FeH, 1.244$\mu m$ and 1.253$\mu m$ KI absorption lines. These lines are weaker in young, low-gravity objects than in older field dwarfs. The HPI index was initially defined for the spectral typing of young objects ($>$M5), but as it traces the shape of the H-band, it also presents a gravity-sensitive behaviour for these later spectral types. The TLI-g index also traces the H-band broadband behaviour using two narrow bands centered at 1.56 and 1.62 microns. It more clearly separates young, mid and field sources and is sensitive over a larger range of spectral types (M0-L3). 

The majority of our objects have indices consistent with being young. Looking at the TLI-g index, there are 5 sources (id: 1, 2, 3, 35, 40) that could be field stars or brown dwarfs. From visual inspection and comparison with field templates we determine that the spectrum of source (id 1) resembles that of a field dwarf. We determine that sources (2, 3, 35 and 40) are young as their spectra better fit the young templates and their other spectral indices are consistent with young objects. The signal-to-noise in the HK band is not sufficient to reliably calculate the TLI-g index for two of our spectra (id 55 and 56), so they have been excluded from the bottom two plots in Fig. \ref{fig:Gravity}.

For our 13 brown dwarfs we also derived the FeH$_J$, KI$_J$ gravity score from \citet{Allers2013}, as this is defined for SpT $\geq M6$. We find 12 of them to have a VL-G classification and one has a INT-G classification (see Table \ref{tab:Big}), where VL-G: very-low gravity and INT-G: intermediate gravity. The brown dwarf with INT-G classification (id 52), has TLI-g and HPI indices consistent with being young. This provides reassurance that these brown dwarfs are young cluster members.



\newpage
\begin{table*}
\scriptsize
\caption{Data table for all KMOS targets. Objects are ordered according the KMOS field and RA; horizontal lines separate the three fields. `X' indicates that a source meets the corresponding selection criteria from \citet{Pearson20}, where \textit{CMD} is the colour-magnitude diagram cut, \textit{VAR} is variability, \textit{KIN} is Gaia kinematics consistent with the locus of NGC 2264 and errors of $<30\%$, \textit{HIGHEX} is located in a region of high extinction. In the \textit{DISC} column, `T' indicates an infrared excess consistent with a thick disc, `A' an anemic disc. \textit{BDC} indicates that a source was selected as a brown dwarf candidate. `-' indicates no data. g-score is the gravity score from \citet{Allers2013}, defined for sources $\geq M6$, where VL-G: very-low gravity and INT-G: intermediate gravity. TLI-g is the index score from Almendros-Abad et al. (in prep.)}
\begin{tabular}{cllccccccccccc}
\hline
Id & RA         & DEC      & CMD & DISC & VAR & KIN & HIGHEX & BDC & SpT  & Av   & Period & g-score & TLI-g  \\
\hline
0  & 100.204818 & 9.620716 & X   & T    & X   &           & X      & X   & -    & -    & -      & -                           & -      \\
1  & 100.208599 & 9.589128 & X   &      &     &           & X      & X   & M5.5 & 2    & -      & -                           & 1.0325 \\
2  & 100.223394 & 9.612833 & X   & T    &     &           & X      &     & M4.5 & 1    & -      & -                           & 1.0456 \\
3  & 100.231364 & 9.644328 & X   & A    &     & X         & X      &     & M5   & 2.5  & -      & -                           & 1.0512 \\
4  & 100.232424 & 9.60172  & X   & T    & X   &           & X      & X   & M4.5 & 18   & -      & -                           & 0.9801 \\
5  & 100.234162 & 9.539768 & X   &      &     &           &        &     & -    & -    & -      & -                           & -      \\
6  & 100.235042 & 9.579001 & X   &      &     &           & X      & X   & M5   & 2    & -      & -                           & 1.0033 \\
7  & 100.236381 & 9.636192 & X   & T    &     &           & X      & X   & M7.5 & 1    & -      & VL-G                        & 0.9392 \\
8  & 100.239426 & 9.587459 & X   &      &     &           & X      & X   & M6   & 3.5  & -      & VL-G                        & 0.9934 \\
9  & 100.24408  & 9.539788 & X   & A    &     &           &        &     & -    & -    & -      & -                           & -      \\
10 & 100.24705  & 9.5662   & X   & A    &     &           & X      &     & M4.5 & 10   & -      & -                           & 0.9795 \\
11 & 100.248405 & 9.616027 & X   & T    &     & X         & X      & X   & M6.5 & 2.5  & 6.44   & VL-G                        & 0.9317 \\
12 & 100.248774 & 9.572889 & X   & T    &     &           & X      & X   & M6.5 & 6    & -      & VL-G                        & 0.9471 \\
13 & 100.249861 & 9.583639 & X   & T    & X   &           & X      &     & -    & -    & -      & -                           & -      \\
14 & 100.25098  & 9.608719 & X   & T    & X   &           & X      &     & M3.5 & 13.5 & -      & -                           & 1.0162 \\
15 & 100.252595 & 9.602863 & X   & T    &     &           & X      &     & M2   & 10.5 & -      & -                           & 0.9862 \\
16 & 100.255925 & 9.647424 & X   & A    &     &           & X      &     & -    & -    & -      & -                           & -      \\
17 & 100.25644  & 9.575863 & X   &      & X   &           & X      &     & -    & -    & 3.68   & -                           & -      \\
18 & 100.257589 & 9.62051  & X   & A    &     &           & X      &     & -    & -    & -      & -                           & -      \\
19 & 100.273353 & 9.565328 & X   & T    &     &           & X      & X   & M5   & 16   & -      & -                           & 1.0007 \\
20 & 100.278984 & 9.598884 & X   &      &     &           & X      & X   & M5.5 & 1    & 0.77   & -                           & 0.9715 \\
21 & 100.280737 & 9.589137 & X   & T    & X   &           & X      & X   & M5   & 10   & -      & -                           & 1.0016 \\
22 & 100.296046 & 9.570143 & X   & T    &     &           & X      &     & M3   & 8.5  & 3.03   & -                           & 0.9514 \\
23 & 100.27607  & 9.4946   & X   &      &     &           &        &     & M6.5 & 11   & -      & VL-G                        & 0.9605 \\
\hline
24 & 100.27932  & 9.510568 & X   & T    &     &           & X      &     & M2.5 & 16   & -      & -                           & 0.9855 \\
25 & 100.28157  & 9.500175 & X   & T    &     &           &        &     & -    & -    & -      & -                           & -      \\
26 & 100.28566  & 9.518336 & X   & T    & X   &           & X      & X   & M4.5 & 10   & -      & -                           & 1.0031 \\
27 & 100.29305  & 9.476    & X   &      & X   & X         & X      &     & -    & -    & -      & -                           & -      \\
28 & 100.29912  & 9.520116 & X   &      & X   &           & X      & X   & M6.5 & 1.5  & -      & VL-G                        & 0.9313 \\
29 & 100.29983  & 9.485867 & X   &      & X   &           & X      &     & -    & -    & -      & -                           & -      \\
30 & 100.30544  & 9.503409 & X   & T    & X   &           &        &     & M2   & 14   & -      & -                           & 0.9891 \\
31 & 100.3058   & 9.506637 & X   & T    & X   &           & X      & X   & M6.5 & 7    & -      & VL-G                        & 0.9621 \\
32 & 100.31614  & 9.51042  & X   & T    &     &           & X      & X   & M6   & 8    & -      & VL-G                        & 0.9365 \\
33 & 100.31892  & 9.563846 & X   &      &     &           & X      &     & -    & -    & -      & -                           & -      \\
34 & 100.31991  & 9.506275 & X   & T    & X   &           & X      &     & M4.5 & 13   & 3.03   & -                           & 0.9706 \\
35 & 100.32141  & 9.484752 & X   &      & X   &           & X      & X   & M5.5 & 2.5  & 2.15   & -                           & 1.0195 \\
36 & 100.3332   & 9.47936  & X   &      &     &           & X      & X   & M6.5 & 1    & -      & VL-G                        & 0.946  \\
37 & 100.33631  & 9.528033 & X   & A    &     &           & X      &     & -    & -    & -      & -                           & -      \\
38 & 100.3444   & 9.512021 & X   &      &     &           & X      &     & -    & -    & -      & -                           & -      \\
39 & 100.34607  & 9.551159 & X   & A    & X   &           &        &     & -    & -    & -      & -                           & -      \\
40 & 100.34704  & 9.541756 & X   & T    &     &           &        & X   & M7   & 2.5  & 3.97   & VL-G                        & 0.9841 \\
41 & 100.34815  & 9.503518 & X   &      &     &           & X      &     & -    & -    & -      & -                           & -      \\
42 & 100.35587  & 9.547427 & X   & T    &     &           &        & X   & -    & -    & -      & -                           & -      \\
43 & 100.36152  & 9.465027 & X   & A    &     &           &        &     & -    & -    & -      & -                           & -      \\
44 & 100.36363  & 9.468747 & X   & A    & X   &           &        &     & -    & -    & -      & -                           & -      \\
45 & 100.37045  & 9.528975 & X   &      &     &           &        &     & -    & -    & -      & -                           & -      \\
\hline
47 & 100.25153  & 9.817421 & X   &      &     &           &        &     & -    & -    & -      & -                           & -      \\
48 & 100.2645   & 9.884649 & X   & A    &     &           &        &     & -    & -    & -      & -                           & -      \\
49 & 100.26457  & 9.877877 & X   &      &     &           &        &     & -    & -    & -      & -                           & -      \\
50 & 100.27362  & 9.895738 & X   & A    & X   &           &        &     & -    & -    & -      & -                           & -      \\
51 & 100.27504  & 9.812569 & X   & A    &     &           & X      &     & -    & -    & -      & -                           & -      \\
52 & 100.27515  & 9.823133 & X   & T    &     &           &        & X   & M7   & 1    & -      & INT-G                       & - \\
53 & 100.2778   & 9.8589   & X   & A    &     &           & X      &     & -    & -    & -      & -                           & -      \\
54 & 100.27865  & 9.848363 & X   &      &     & X         &        &     & -    & -    & -      & -                           & -      \\
55 & 100.286    & 9.89631  & X   &      & X   & X         &        & X   & M6.5 & 0    & -      & VL-G                        & 1      \\
56 & 100.2863   & 9.890197 & X   &      &     &           & X      & X   & M8   & 0.5  & -      & VL-G                        & - \\
57 & 100.2903   & 9.886707 & X   &      & X   &           & X      &     & -    & -    & -      & -                           & -      \\
58 & 100.32122  & 9.825583 & X   &      &     &           &        &     & -    & -    & -      & -                           & -      \\
59 & 100.32194  & 9.870709 & X   &      &     &           &        &     & -    & -    & -      & -                           & -      \\
60 & 100.32544  & 9.799609 & X   & A    &     &           &        &     & -    & -    & -      & -                           & -      \\
61 & 100.32662  & 9.859099 & X   &      &     &           &        &     & -    & -    & -      & -                           & -      \\
62 & 100.32754  & 9.902132 & X   &      &     &           &        &     & M3   & 1.5  & -      & -                           & 1.0159 \\
63 & 100.32769  & 9.88348  & X   &      & X   &           &        & X   & -    & -    & 1.26   & -                           & -      \\
64 & 100.32773  & 9.817334 & X   & T    &     &           &        & X   & M5.5 & 0    & -      & -                           & 0.9899 \\
65 & 100.32895  & 9.81071  & X   & A    &     &           &        &     & -    & -    & 1.66   & -                           & -      \\
66 & 100.3329   & 9.860489 & X   &      &     &           &        &     & -    & -    & -      & -                           & -      \\
67 & 100.33944  & 9.883792 & X   & A    &     &           &        & X   & M3.5 & 1    & -      & -                           & 1.0158 \\
68 & 100.34215  & 9.889519 & X   &      &     &           &        &     & -    & -    & -      & -                           & -      \\
\hline
\label{tab:Big}
\end{tabular}
\end{table*}

\section{Properties of confirmed brown dwarfs}

We spectroscopically confirm 13 sources as young brown dwarfs in NGC 2264. A further 19 are identified as young early-mid M-type stars from our spectroscopy. In the following, we briefly discuss the properties of these samples. In Fig. \ref{fig:3pan} we show the colours of these samples compared to the full population of candidates identified in \citet{Pearson20}.

All 32 objects are optically faint, with I-band magnitudes ranging from 17.5 to 22. They extend the known YSO population in NGC 2264 with spectra towards fainter magnitudes. From our spectral fitting, the optical extinction towards these sources ranges from $A_V$ of 0 to 18. The fact that the sources cover a wide range in extinction is also apparent from the near-infrared colours. While the low-mass stars and brown dwarfs are well mixed in the (I-J,I) colour magnitude diagram and to some extent the (I-K, J-H) colour-colour plot (Fig. \ref{fig:IKJH}), they separate in the near-infrared colour-colour space, with all 13 brown dwarfs being located below M6 the extinction vector within errorbars in the (J-H, H-K) colour-colour plot (Fig. \ref{fig:3pan} middle). This bolsters our initial selection of brown dwarf candidates based on these colours.

With very few exceptions, the confirmed sources with Spitzer data show excess emission in the mid-infrared, as inferred from the Spitzer/IRAC slope. This is again a feature of our selection process; we used infrared colour excess as criterion for youth. In particular, of the 8 confirmed brown dwarfs with Spitzer data, 7 exhibit an infrared excess indicative of a disc.


Of the confirmed brown dwarfs, 6/13 have Gaia DR2 kinematics. All six of these brown dwarfs show Gaia parallax and proper motions consistent with the known members of NGC 2264. As they are optically very faint sources, they are close to Gaia's detection limit and as such the associated error bars in their kinematics are very large.

The optical extinction values for the confirmed brown dwarfs range between $A_V$ of 0 and 11. Extinction does not appear to affect the brown dwarf confirmation rate as brown dwarfs have been identified across the range of extinctions. Before the extinction was more accurately determined through spectral fitting, 9/13 of the confirmed brown dwarfs were identified as being located in high extinction regions using a NIR extinction map. It is worth noting that our three chosen KMOS fields cover known high extinction areas, therefore finding brown dwarfs with high $A_V$ is not unexpected.

\begin{figure}
	\centering
  	\includegraphics[width=.45\textwidth]{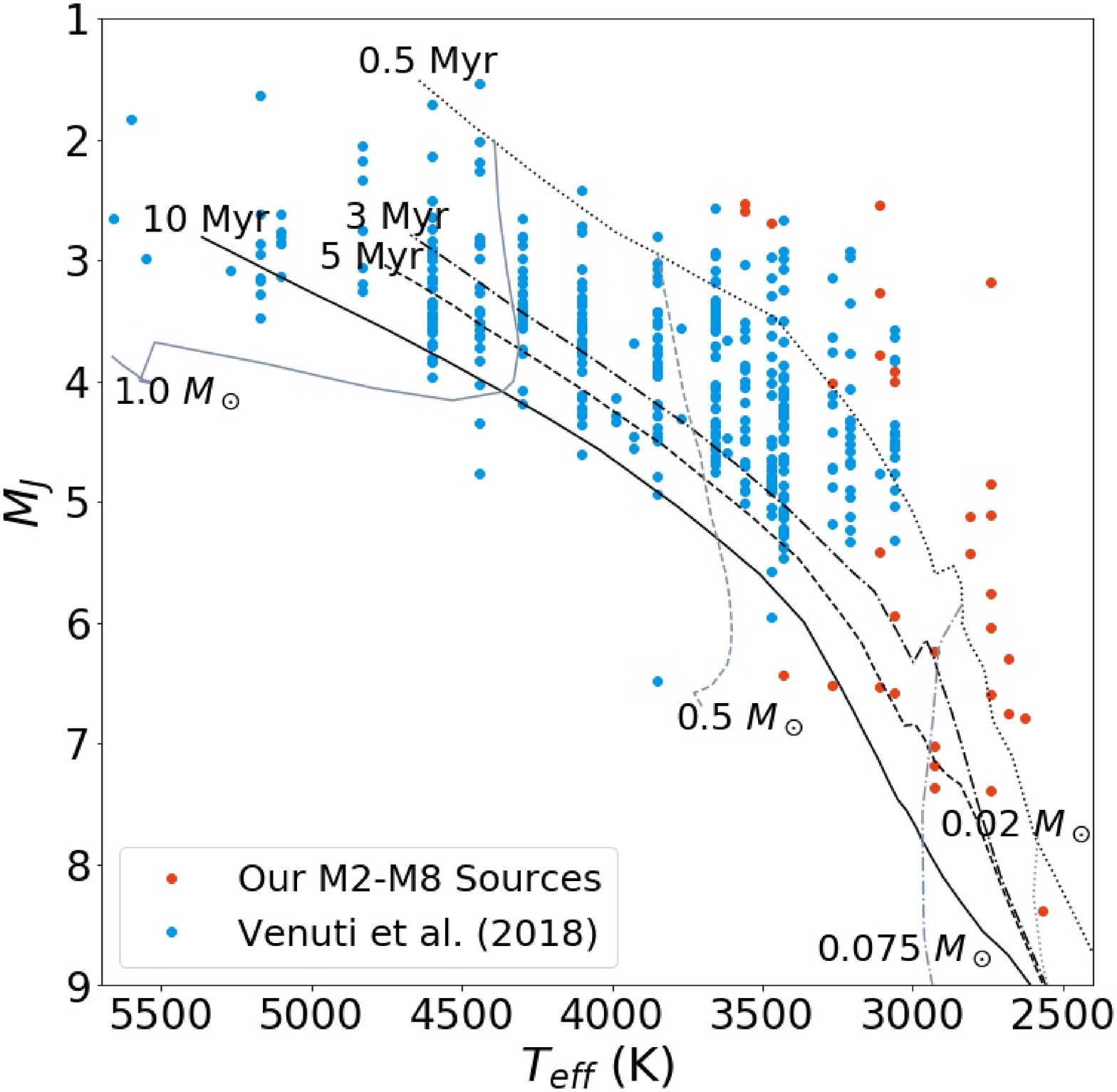}
	\caption{The absolute J-band magnitude vs effective temperature for the 32 low-mass sources identified in this study (red) as well as the \citet{Venuti2018} sample of spectroscopically confirmed cluster members (blue). Isochrones for 0.5, 3, 5 \& 10 Myr are plotted in black, mass tracks for 1$M_\odot$, 0.5$M_\odot$, 0.075$M_\odot$ and 0.02$M_\odot$ are plotted in grey \citet{Baraffe}.} 
	\label{fig:MJ_Teff}
\end{figure}

In Fig. \ref{fig:MJ_Teff} we show the HR diagram, with absolute J-band magnitude plotted against effective temperature (Teff), for the 32 young sources identified in this study, as well as the \citet{Venuti2018} sample of spectroscopically confirmed cluster members. Isochrones for 0.5, 3, 5 and 10 Myr are plotted in black, evolutionary tracks for 1$M_\odot$, 0.5$M_\odot$, 0.075$M_\odot$ and 0.02$M_\odot$ in grey \citep{Baraffe}.

The absolute J-band magnitudes were calculated using the extinctions derived from spectral fitting and the distance of 719\,pc \citep{2264dist} for the cluster. The effective temperatures were derived from the spectral type using the tabulated intrinsic colours and temperatures of 5-30 Myr old pre-main-sequence stars \citep{PandM}. The binning in Teff, visible in Fig. \ref{fig:MJ_Teff}, is due to the conversion from spectral type to Teff, as spectral type is a discrete sequence. The typical error for the Teff is $\pm$ 160K, which comes from the associated error for the spectral type being $\pm$ 1 sub type.

\begin{figure}
	\centering
  	\includegraphics[width=.45\textwidth]{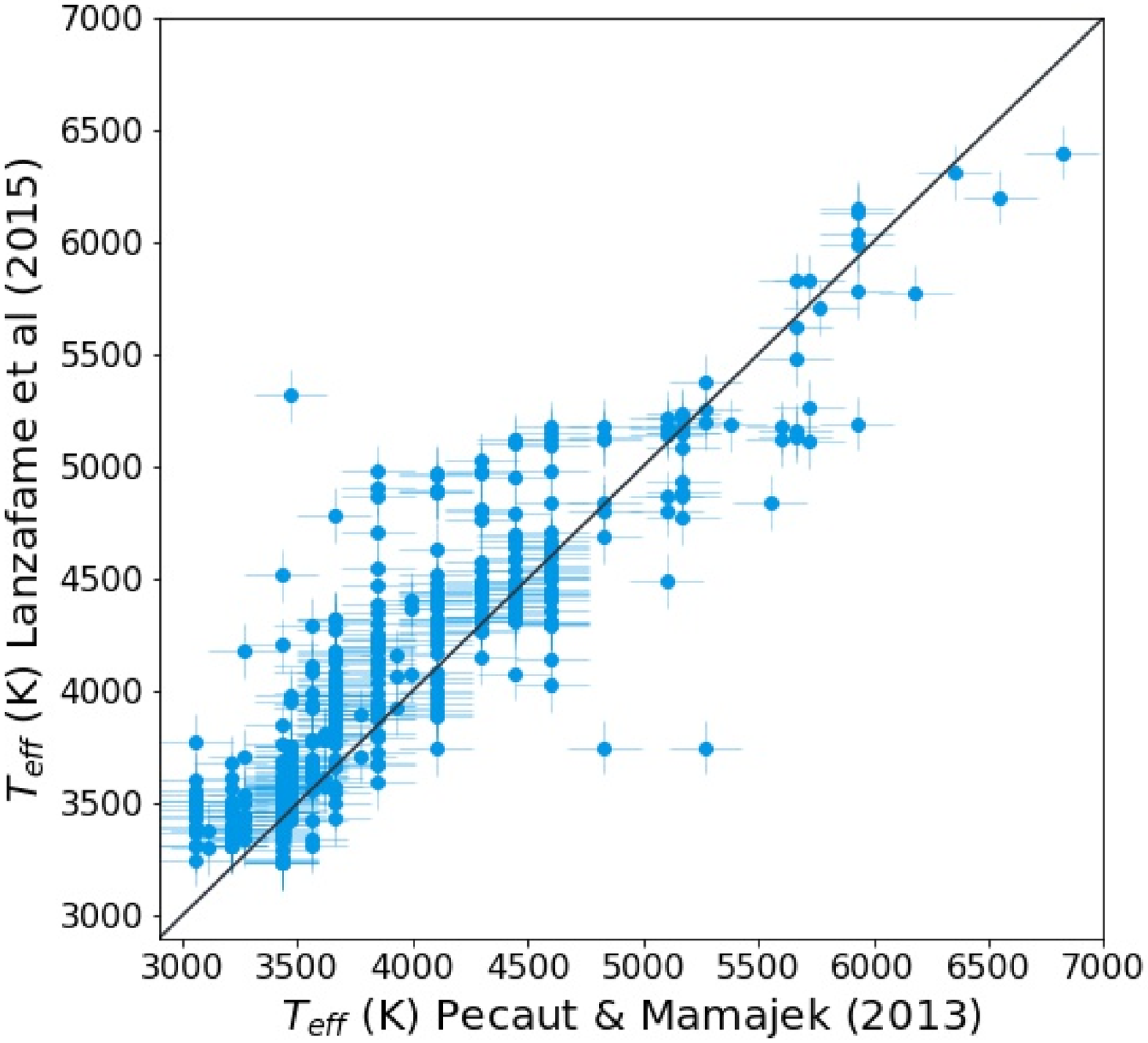}
	\caption{A comparison of the effective temperature derived by \citet{Lanzafame2015} (used by \citet{Venuti2018}) to the effective temperature from derived from the spectral type using a series of intrinsic colours and temperatures of 5-30 Myr old pre-main-sequence stars \citep{PandM}.} 
	\label{fig:TeffComp}
\end{figure}

In Fig. \ref{fig:TeffComp} we compare the Teff derived by \citet{Lanzafame2015} (used by \citet{Venuti2018}) to the Teff calculated from the given spectral type. \citet{Lanzafame2015} derive Teff by fitting a series of broadened and \textit{vieled} template spectra to their target spectra. The template spectra that most closely reproduce the target spectrum are selected, and their weighted average Teff assigned to the target star. Compared to \citet{Lanzafame2015}, the spectral type to Teff conversion underpredicts Teff for the cooler stars and overpredicts Teff for the hotter stars. Overall the two estimates are in reasonable agreement, within the uncertainties of $\pm 120$\,K for the values from \citet{Lanzafame2015} and $\pm$ 160K for the spectral type conversion. We decided to use the spectral type to Teff conversion for all sources to have a consistent method across both samples. 

In the HR diagram, our sample of 32 new sources follows on from the low-mass end of the Venuti sample, with some overlap in the 3100-3500K range. The lowest mass brown dwarf that we confirm has a mass of around $0.02\,M_{\odot}$. In comparison with the isochrones, the ages range from $<0.5$ to $>5$\,Myr. By and large, the age distribution in our sample is similar to the more massive stars in the cluster \citep{Sung2010, Venuti2018, nony2020}. We note that more than half of the brown dwarfs and very low mass stars identified in this study are comparatively very young, according to the isochrones, with an estimated age of $<0.5$\,Myr. This is not unexpected as we specifically targeted sources with discs in high extinction areas, i.e. we expect a bias towards very young objects, but may also be an systematic artefact due to binarity \citep{Sullivan_2021}.

\section{The brown dwarf population in NGC 2264}

\subsection{The total number of brown dwarfs}

In this subsection we aim to estimate the total number of brown dwarfs in this cluster. 

\begin{figure}
	\centering
  	\includegraphics[width=.45\textwidth]{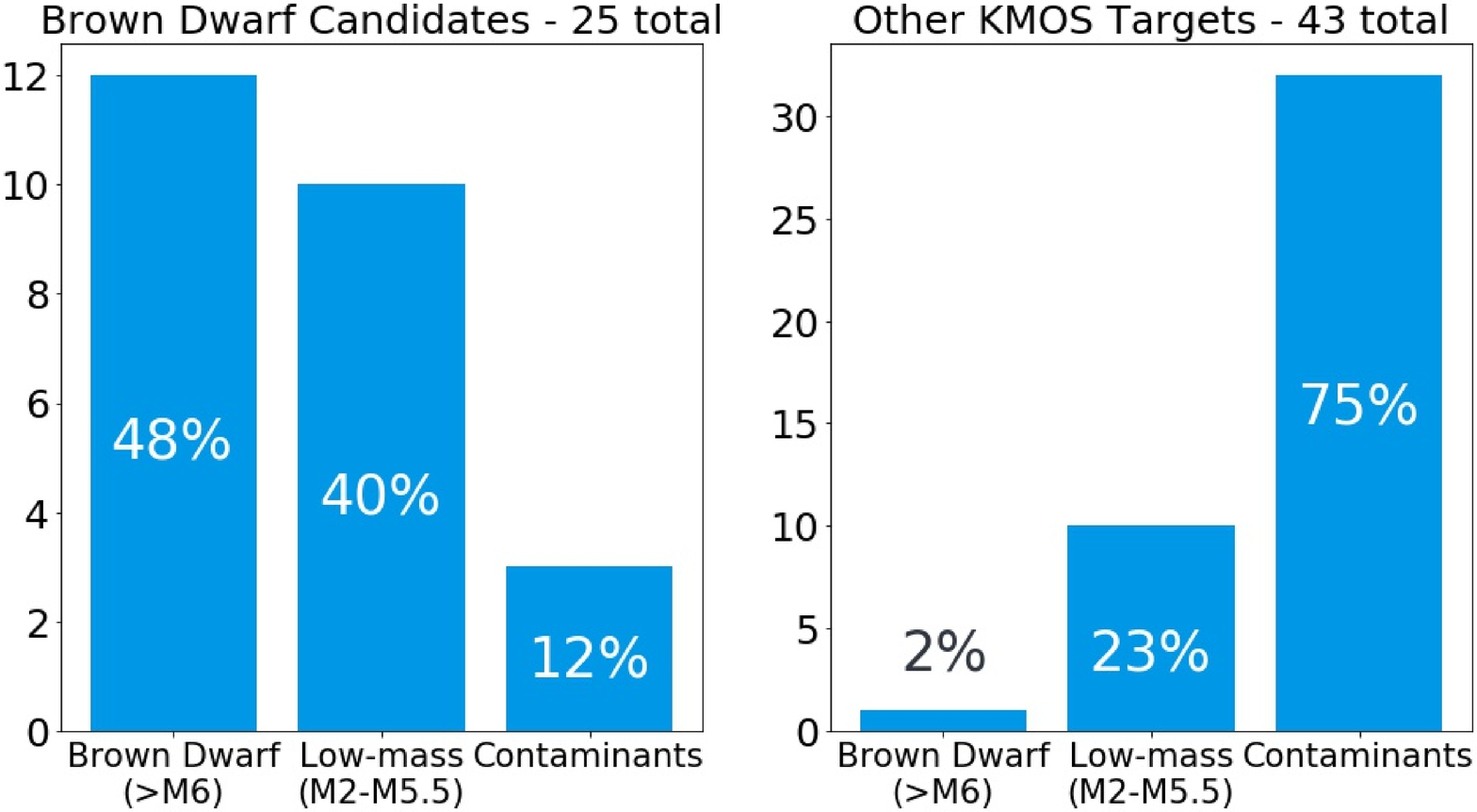}
	\caption{The results of the spectral classifications for the 25 brown dwarf candidates (left) and the remaining 43 targets (right), split in to three categories: brown dwarfs ($\ge M6$), early-mid M stars (M2-M5.5) and contaminants.} 
	\label{fig:SpT_re}
\end{figure}

First, we check the success rate of our follow-up spectroscopy. We obtained near infrared spectra for a total of 68 sources, 25 of these sources were identified as brown dwarf candidates in \citet{Pearson20}. The success rate estimates for this sample are shown in Fig. \ref{fig:SpT_re} (left). 12/25 (48\%) were confirmed to be brown dwarfs ($\ge M6$), 10/25 (40\%) were found to be early-mid M stars (M2-M5.5), and 3/25 (12\%) were rejected as contaminants. The contaminants were identified as having a spectrum that is well approximated by a reddened blackbody, but gave a poor fit when compared with the young low-mass template spectra. These are likely either embedded young stars with spectral types significantly earlier than M0, or reddened background objects. 


The distribution of the spectral types determined for the remaining 43 sources is also illustrated in Fig. \ref{fig:SpT_re} (right). 1/43 (2\%) were found to be substellar, 10/43 (23\%) were found to be early-mid M stars, and 32/43 (74\%) were identified as background/contaminants.

With the help of the follow-up for a limited number of candidates, we can estimate the likely number of all brown dwarfs in NGC 2264. 
There are 429 brown dwarf candidates identified in \citet{Pearson20}. From our spectroscopy success rate we expect about half ($\sim 200$) of these to be confirmed as bona fide brown dwarfs. This will likely underestimate the total number of brown dwarfs in NGC 2264 as the identified brown dwarf candidates are biased towards variable and disc bearing sources. These sources are typically the youngest sources in the cluster, meaning that many of the older more evolved brown dwarfs are missing from the candidate list. This is very clearly seen in Fig. \ref{fig:MJ_Teff}, where the majority of the sources identified in this study lie above the 0.5 Myr isochrone. For this reason we take 200 as the lower bound for the total number of brown dwarfs in NGC 2264.

Most of the confirmed brown dwarfs show infrared excess due to a disc, but overall we expect that only about one third of the total substellar population NGC 2264 will host a disc. This assumption is based on the disc fraction for the stars. We expect the disc fraction for stars and brown dwarfs in NGC 2264 to be approximately the same, as this is seen in other young regions \citep{dawson2013}. \citet{Venuti2018} spectroscopically identified 655 cluster members between $0.2$ and $1.8M_\odot$. Although this sample is incomplete in terms of membership consensus and will miss the more embedded sources, it is a very clean sample of stars, with low contamination. For the 574/655 sources in this sample that have Spitzer data, 197 (34\%) exhibit infrared excess indicative of a disc. Using that disc fraction, we estimate that up to two thirds of brown dwarfs without discs may be missed by our selection of objects for spectroscopy. That means, a conservative upper bound  for the total number of brown dwarfs in NGC 2264 is $\sim 600$.

In summary, we expect that NGC 2264 hosts between 200-600 brown dwarfs with masses down to 0.02$\,M_{\odot}$.

The brown dwarf abundance in other young star forming regions is typically 1:2 - 1:5 (brown dwarfs:stars) for masses down to 0.03$\,M_{\odot}$, i.e 1 brown dwarf for every 2 - 5 stars \citep{kora2017, kora2019}. All estimates of that quantity for young star forming regions are so far in agreement within the considerable errorbars. For NGC 2264 with an estimated stellar population of $\sim1500$ \citep{Teixeira2012, Rapson2014, Venuti2018}, this corresponds to a total of $\sim300 - 750$ brown dwarfs. Our estimate of $200-600$ falls roughly in the middle of this range. Hence, the abundance of brown dwarfs in NGC 2264 found by our survey is in line with estimates for other regions thus far.

\subsection{The substellar mass function}
In this section we use our confirmed brown dwarfs and remaining young and brown dwarf candidates to explore the substeller mass function (MF) in NGC 2264. We limit the spatial extent of this analysis to the area covered by the three KMOS fields (see Fig. \ref{fig:TarSpatial}), as within these fields we have information about the typical extinction from our KMOS spectra.

For the 13 brown dwarfs we have confirmed within these fields, we deredden the J band photometry using the extinction determined through template fitting and estimate the mass using the BHAC15 1 Myr isochrone \citep{Baraffe} and a distance of 719\,pc \citep{2264dist}.

\begin{figure}
	\centering
  	\includegraphics[width=.45\textwidth]{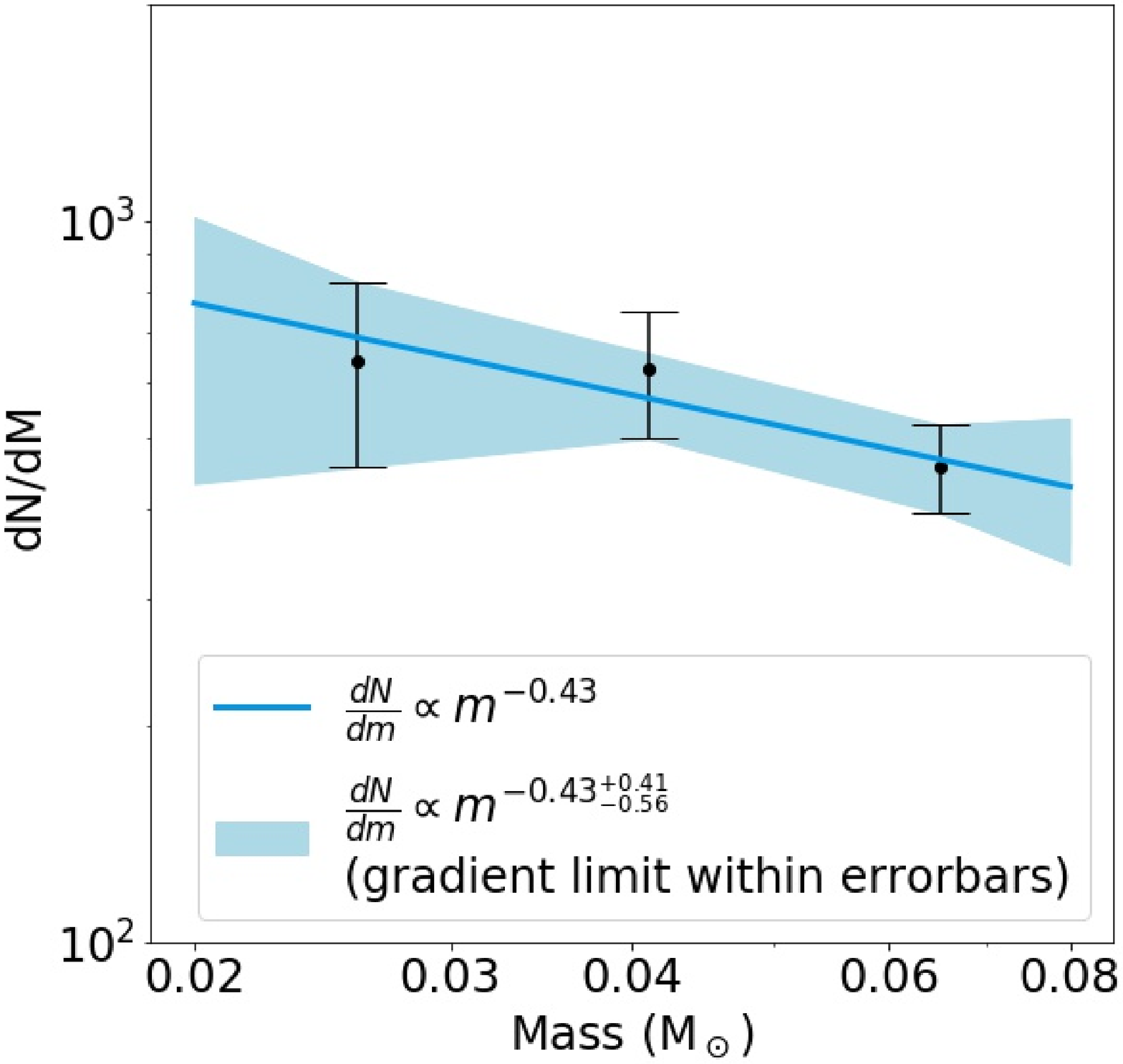}
	\caption{The substeller MF for the three KMOS fields in NGC 2264 (see Fig. \ref{fig:TarSpatial}), represented with three equally sized log-mass bins, in the range 0.02M$_\odot$ - 0.08M$_\odot$. The best fitting power law of the form $\frac{dN}{dm} \propto m^{-\alpha}$ with $\alpha = 0.43$ is shown by the dark blue line. The light blue shading indicates the range of acceptable fits within the errorbars for the three bins, with $\alpha = 0.43^{+0.41}_{-0.56}$} 
	\label{fig:IMF}
\end{figure}

In addition to the confirmed brown dwarfs, there are 27 remaining brown dwarf candidates, and a further 43 sources with indicators of youth (VAR, HIGHVAR, DISC, KIN, HIGHEX) that are within the three fields and are brighter than 18 magnitude in J, the limit of our KMOS observations. For these 70 sources we do not have spectra and cannot estimate masses in the same way as described above. Instead, we assume that the `success rate' of our spectroscopic confirmation would also apply to this sample, i.e. if we were to take spectra for all of them, we would confirm 48\% of the brown dwarf candidates and 2\% of the sources with indicators of youth, as brown dwarfs (Fig. \ref{fig:SpT_re}). We also assume that the extinctions determined from spectroscopy are representative for the entire sample. With these two assumptions, the masses are then estimated as follows,

First, we randomly pick a subsample from these 70 sources using our success rates for the different samples. We randomly assign an extinction to each source drawn from the 32 values for $A_V$ determined by spectroscopy. We then deredden the J band photometry and calculate the mass, again using the BHAC15 1 Myr isochrone. We repeat this process 1000 times and report the average number of sources in three equally sized log-mass bins, in the range 0.02M$_\odot$ - 0.08M$_\odot$. We take the standard deviation from the iterations as the error for each bin. We then combine this with the mass distribution of the spectroscopically confirmed brown dwarfs to determine the substeller MF for NGC 2264 (Fig. \ref{fig:IMF}).

We fit a power law of the form $\frac{dN}{dm} \propto m^{-\alpha}$ and find the best fit is given by $\alpha = 0.43$. We define the error for $\alpha$ by the range of acceptable fits within the errorbars for the three bins, which gives $\alpha = 0.43^{+0.41}_{-0.56}$.

Our best fitting slope $\alpha$ is consistent with the slopes determined for the brown dwarf regime in nearby young clusters (see \citet{kora2017} for a compilation), for example: IC 348 $\alpha = 0.7\pm0.4$ \citep{Alves_de_Oliveira_2013}, $\rho$ Oph $\alpha = 0.7\pm0.3$ \citep{Alves_de_Oliveira_2012}, Upper Scorpius $\alpha = 0.45\pm0.11$ \citep{Lodieu_2013}. At face value, this points towards a universal substellar IMF across star-forming environments.
It is noteworthy that the substellar $\alpha$ we estimate for NGC 2264 is at the lower end of the range of the values quoted above. The same is true for RCW 38, a very massive, very young cluster, which has  $\alpha = 0.42\pm0.18$ for the mass range 0.02-0.2M$_\odot$ \citep{kora2017}. Whether or not this is a trend that indicates how environment affects the substellar MF remains to be tested in other clusters.





\section{Summary and outlook}

We have used near-infrared spectra from KMOS to perform spectroscopic follow-up observations of 68 red, faint candidates from our multi-epoch, multi-wavelength survey of NGC 2264 \citep{Pearson20}. By fitting with spectral standards we measure spectral types and extinction values for 32 young M-type cluster members. We also use spectral indices for an independent spectral type determination and an estimate of the uncertainty.

We confirm 13 brown dwarfs in NGC 2264, the first spectroscopically confirmed brown dwarfs in this cluster. The confirmation rate for brown dwarf candidates identified in our previous survey  was found to be 48\%. A further 19 objects are found to be young early-mid M-type members of NGC 2264, with spectral types of M2 - M5.5. The spectral types for the brown dwarfs range between M6 and M8, corresponding to masses between 0.02 and 0.08\,$M_{\odot}$. Comparing with isochrones, more than half of the brown dwarfs and very low mass stars identified in this study are comparatively very young, with an estimated age of $<0.5$\,Myr. The remaining sources show an age spread of 0.5 to 5\,Myr, comparable to the age of the young stars in the cluster.

Based on the success rate of our spectroscopic follow-up, we estimate that NGC 2264 hosts between 200 and 600 brown dwarfs in total (in the given mass range). This is consistent with the star-to-brown dwarf ratio found in many other star forming regions. We also determine the slope of the substellar mass function as $\alpha = 0.43$ (with large uncertainty), consistent with those measured for other young clusters. This points to a uniform substellar mass function across all star forming environments.


We have shown that while more distant than the nearest star forming regions, NGC 2264 is still accessible for detailed spectroscopic follow-up well into the substellar regime. As a very rich and massive cluster ($\sim1500$ stars, $200-600$ brown dwarfs) it would be ideal for studies that benefit from a large sample size of brown dwarfs, such as environmental factors on brown dwarf formation efficiency and the substellar mass function. Its relatively compact size on the sky also makes it an ideal target for further followup with JWST.

\section*{Acknowledgements}
Based on observations collected at the European Southern Observatory under ESO programme 0104.C-0105(A). We acknowledge support from STFC through grant number ST/R000824/1. K.M. acknowledges funding by the Science and Technology Foundation of Portugal (FCT), grants No. IF/00194/2015, PTDC/FIS-AST/28731/2017 and UIDB/00099/2020. This work is based [in part] on observations made with the Spitzer Space Telescope, which was operated by the Jet Propulsion Laboratory, California Institute of Technology under a contract with NASA. This work has made use of data from the European Space Agency (ESA) mission
{\it Gaia} (\url{https://www.cosmos.esa.int/gaia}), processed by the {\it Gaia}
Data Processing and Analysis Consortium (DPAC,
\url{https://www.cosmos.esa.int/web/gaia/dpac/consortium}). Funding for the DPAC
has been provided by national institutions, in particular the institutions
participating in the {\it Gaia} Multilateral Agreement.
This research has benefitted from the Montreal Brown Dwarf and Exoplanet Spectral Library, maintained by Jonathan Gagné.

\section*{Data Availability}
This article is based on observations collected at the European Southern Observatory under ESO programme 0104.C-0105(A). The data underlying the research results are shown in Table \ref{tab:Big}.




\bibliographystyle{mnras}
\bibliography{Bib.bib} 







\bsp	
\label{lastpage}
\end{document}